\documentclass[preprintnumbers,amsmath,amssymb]{revtex4}
\usepackage{bm}
\usepackage{braket}
\usepackage{mathrsfs}
\usepackage{graphicx}
\usepackage[caption=false]{subfig}
\usepackage{xcolor}
\usepackage{float}
\usepackage{booktabs}
\usepackage{tabularx}
\usepackage{multirow}
\usepackage{epstopdf}
\usepackage[unicode=true]{hyperref}

\begin{document}

\title{Tight Trade-off Between Internal, Assisted, and External Entanglement}

\author{Limin Gao}
\email{gaoliminabc@163.com}
\affiliation{School of Mathematics and Science, Hebei GEO University, Shijiazhuang 052161, China}
\affiliation{Hebei Province Key Laboratory of Intelligent Sensing and Data Processing for Geo-environment, Hebei GEO University, Shijiazhuang 052161, China}
\author{Chenxiao Wang}
\affiliation{School of Mathematics and Science, Hebei GEO University, Shijiazhuang 052161, China}

\begin{abstract}
We derive a tight and saturable monogamy relation for three-qubit pure states that bounds the sum of concurrence and concurrence of assistance by the entanglement with an external qubit. The bound decreases strictly with increasing external entanglement, establishing a precise trade-off between internal and environment-induced entanglement. Equivalent formulations in terms of negativity and its convex-roof extensions follow. Our result provides a unified and quantitative constraint on entanglement distribution in open multipartite quantum systems.
\end{abstract}

\maketitle

\section{Introduction}

Quantum entanglement cannot be freely shared among multiple parties, a restriction formalized by monogamy relations \cite{12,2a}. Such constraints are central to quantum information science, with implications ranging from secure communication to many-body physics.

Most known monogamy relations are derived for effectively closed systems \cite{10b,14b,15b,20b,28b,13b,34b}. In realistic settings, however, subsystems are unavoidably coupled to external degrees of freedom, such that entanglement with the environment competes with that shared internally. Although this competition has been examined in several works \cite{38b,39b,40b,41b}, existing approaches either treat internal and external correlations separately or fail to yield tight bounds. A unified and tight constraint capturing the interplay among internal, assisted, and external entanglement remains lacking.

In this Letter, we establish such a relation for three-qubit pure states. Extending previous approaches, we show that for a two-qubit subsystem $AB$ of a pure state $|\psi\rangle_{ABC}$, the sum of its concurrence and concurrence of assistance is tightly bounded by the entanglement between $AB$ and the remaining qubit $C$. The bound is saturable and decreases monotonically with increasing external entanglement, revealing a precise trade-off between internal entanglement and entanglement shared with the environment.

Importantly, the same structure extends directly to negativity, convex-roof extended negativity (CREN), and CREN of assistance (CRENoA), providing a unified description across distinct entanglement measures. Our results establish a fundamental limitation on entanglement sharing in open multipartite systems and offer a quantitative tool for analyzing entanglement distribution in realistic quantum architectures.

\section{Monogamy Relations}

For a bipartite pure state $|\psi\rangle_{AB}=\sum_i \sqrt{\lambda_i}\,|ii\rangle$, the concurrence is defined as \cite{12,62}
\begin{equation}
C(|\psi\rangle_{AB})=\sqrt{2\!\left[1-\operatorname{Tr}(\rho_A^2)\right]},
\end{equation}
where $\rho_A=\operatorname{Tr}_B(|\psi\rangle\langle\psi|)$. For a mixed state $\rho_{AB}$, its convex-roof extension is
\begin{equation}
C(\rho_{AB})=\min_{\{p_j,|\psi_j\rangle\}} \sum_j p_j\, C(|\psi_j\rangle),
\end{equation}
whereas the concurrence of assistance is \cite{63}
\begin{equation}
C_a(\rho_{AB})=\max_{\{p_j,|\psi_j\rangle\}} \sum_j p_j\, C(|\psi_j\rangle).
\end{equation}

The negativity is defined by \cite{42,43,45}
\begin{equation}
\mathcal{N}(\rho_{AB})=\|\rho_{AB}^{T_A}\|_1-1,
\end{equation}
and its convex-roof extensions are \cite{46,18}
\begin{equation}
\begin{aligned}
\widetilde{\mathcal{N}}(\rho_{AB}) &= \min_{\{p_k,|\varphi_k\rangle\}} \sum_k p_k\, \mathcal{N}(|\varphi_k\rangle),\\
\widetilde{\mathcal{N}}_a(\rho_{AB}) &= \max_{\{p_k,|\varphi_k\rangle\}} \sum_k p_k\, \mathcal{N}(|\varphi_k\rangle).
\end{aligned}
\end{equation}
These quantities satisfy
\begin{equation}
\mathcal{N}(\rho_{AB})\le \widetilde{\mathcal{N}}(\rho_{AB})\le \widetilde{\mathcal{N}}_a(\rho_{AB}).
\end{equation}
For $2\otimes d$ systems,
\begin{equation}
\mathcal{N}(|\psi\rangle)=\widetilde{\mathcal{N}}(|\psi\rangle)=C(|\psi\rangle),
\end{equation}
and hence
\begin{equation}
\widetilde{\mathcal{N}}(\rho_{AB})=C(\rho_{AB}).
\end{equation}
Similarly, for two-qubit states,
\begin{equation}
\widetilde{\mathcal{N}}_a(\rho_{AB})=C_a(\rho_{AB}).
\end{equation}

We now state the main result.

\textbf{Theorem.}---For any three-qubit pure state $|\psi\rangle_{ABC}$ with
$\rho_{AB}=\operatorname{Tr}_C(|\psi\rangle_{ABC}\langle\psi|)$,
one has
\begin{equation}\label{eq:thm1-conc-1}
C(\rho_{AB})+C_a(\rho_{AB})-\sqrt{1-C^2\!\left(|\psi\rangle_{C|AB}\right)}
\le 1.
\end{equation}
Equivalent formulations are
\begin{equation}\label{eq:thm1-cren-1}
\widetilde{\mathcal{N}}(\rho_{AB})+\widetilde{\mathcal{N}}_a(\rho_{AB})-\sqrt{1-\widetilde{\mathcal{N}}^2\!\left(|\psi\rangle_{C|AB}\right)}
\le 1,
\end{equation}
and
\begin{equation}\label{eq:thm1-neg-1}
\mathcal{N}(\rho_{AB})+\widetilde{\mathcal{N}}_a(\rho_{AB})-\sqrt{1-\mathcal{N}^2\!\left(|\psi\rangle_{C|AB}\right)}
\le 1.
\end{equation}

Equation~(\ref{eq:thm1-conc-1}) yields a sharp trade-off between internal and external entanglement: the maximal allowed value of $C(\rho_{AB})+C_a(\rho_{AB})$ decreases monotonically with $C(|\psi\rangle_{C|AB})$. As shown below, the boundary is exactly attainable. The proof is given in the Supplemental Material. Theorem~1 also recovers the bound of Ref.~\cite{40b} as an immediate consequence.

\textbf{Corollary.}---For any three-qubit state $\rho_{ABC}$ with
$\rho_{AB}=\operatorname{Tr}_C(\rho_{ABC})$,
one has
\begin{equation}
C(\rho_{AB})+\frac{1-\sqrt{1-C^2(\rho_{C|AB})}}{2}\le 1.
\end{equation}
Here $C(\rho_{C|AB})$ denotes the bipartite concurrence across the partition $C|AB$.

\emph{Proof}.---Let
\begin{equation}
g(x)=\frac{1+\sqrt{1-x^2}}{2},
\end{equation}
which is concave and monotonically decreasing on $[0,1]$.

For pure states $\rho_{ABC}=|\psi\rangle\langle\psi|$, the result follows directly from Theorem~1:
\begin{equation}
C(\rho_{AB}) \le g\!\left(C(|\psi\rangle_{C|AB})\right).
\end{equation}

For mixed states, consider an arbitrary decomposition
$\rho_{ABC}=\sum_i p_i |\psi_i\rangle\langle\psi_i|$,
and define $\rho_{AB}^i=\operatorname{Tr}_C(|\psi_i\rangle\langle\psi_i|)$.
Using the convexity of concurrence and Theorem~1 applied to each pure state, we obtain
\begin{align}
C(\rho_{AB})
&\le \sum_i p_i C(\rho_{AB}^i) \\
&\le \sum_i p_i g\!\left(C(|\psi_i\rangle_{C|AB})\right).
\end{align}
By concavity of $g$, this implies
\begin{equation}
C(\rho_{AB}) \le g\!\left(\sum_i p_i C(|\psi_i\rangle_{C|AB})\right).
\end{equation}
Finally, by the convex-roof definition of concurrence,
\begin{equation}
\sum_i p_i C(|\psi_i\rangle_{C|AB}) \ge C(\rho_{C|AB}),
\end{equation}
and since $g$ is decreasing, we conclude
\begin{equation}
C(\rho_{AB}) \le g\!\left(C(\rho_{C|AB})\right),
\end{equation}
which is equivalent to the stated inequality.
\hfill$\square$

\begin{figure}[t]
\centering
\includegraphics[width=0.48\columnwidth]{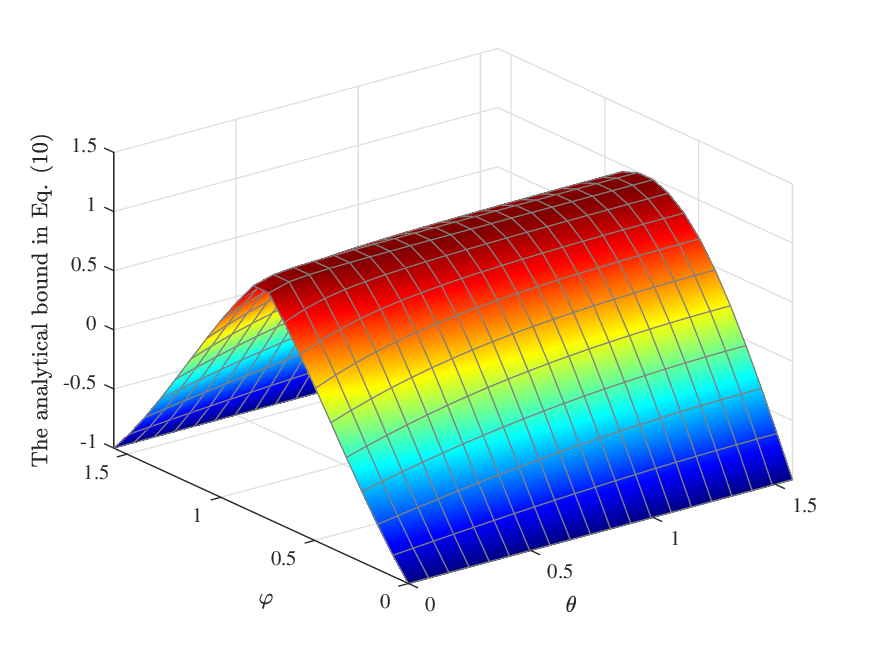}
\includegraphics[width=0.48\columnwidth]{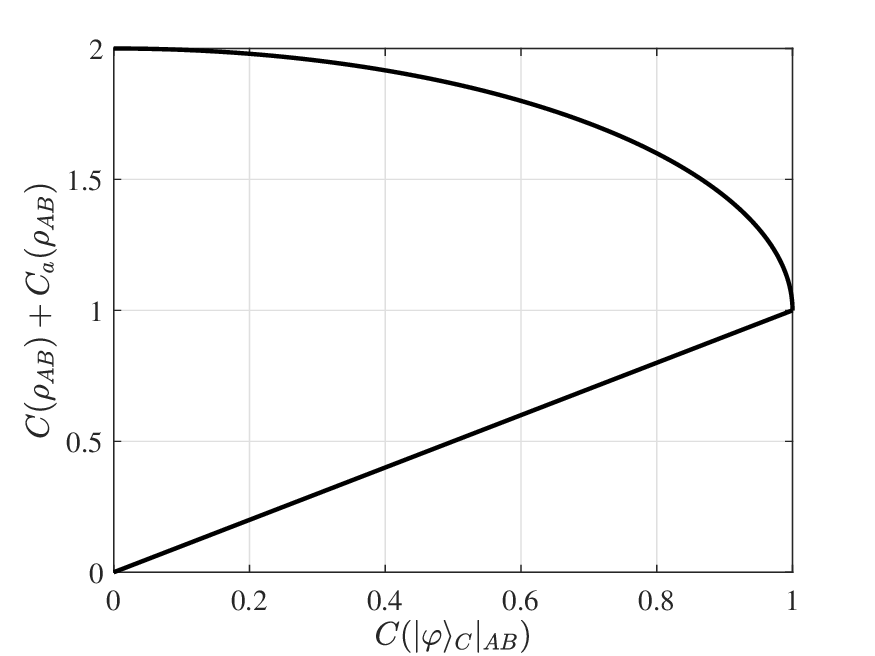}
\caption{Trade-off between internal and external entanglement. (Left) The boundary defined by Eq.~(\ref{eq:thm1-conc-1}) is saturated by the family of states in Eq.~(\ref{eq:example-family}) with $\varphi=\pi/4$, where $\theta$ parametrizes the curve. (Right) The overall constraint on the distribution of internal entanglement $C(\rho_{AB})$, assisted entanglement $C_a(\rho_{AB})$, and external entanglement $C(|\psi\rangle_{C|AB})$.}
\end{figure}

\emph{Example}.---The tightness of Eq.~(\ref{eq:thm1-conc-1}) is demonstrated by the family of states
\begin{equation}\label{eq:example-family}
|\psi\rangle_{ABC} = \sin\varphi\,|000\rangle + \cos\varphi\sin\theta\,|110\rangle + \cos\varphi\cos\theta\,|111\rangle,
\end{equation}
where $\theta,\varphi \in [0, \pi/2]$. For these states, the relevant entanglement measures are
\begin{align}
C(\rho_{AB}) &= \sin 2\varphi \sin\theta, \quad C_a(\rho_{AB}) = \sin 2\varphi, \quad C(|\psi\rangle_{C|AB}) = \sin 2\varphi \cos\theta.
\end{align}
Setting $\varphi = \pi/4$ maximizes the internal correlations, yielding
\begin{equation}
C(\rho_{AB}) + C_a(\rho_{AB}) = 1 + \sin\theta = 1 + \sqrt{1 - C^2(|\psi\rangle_{C|AB})},
\end{equation}
which identically saturates Eq.~(\ref{eq:thm1-conc-1}) for all $\theta \in [0, \pi/2]$. By varying $\theta$, this family maps out the exact boundary of the allowed region, illustrating a strict trade-off: as the internal entanglement grows, the capacity for external entanglement is correspondingly depleted, as shown in Fig.~1.

\section{Conclusion}

We have established a tight monogamy relation that simultaneously constrains internal entanglement, assisted entanglement, and entanglement with an external qubit in three-qubit systems. The bound is saturable and provides a unified characterization of entanglement sharing in open multipartite settings. Because the same structure extends directly to negativity and its convex-roof variants, our result offers a common framework for several standard entanglement measures and may be useful for analyzing entanglement distribution in realistic quantum architectures.

\section*{Supplemental Material}

\renewcommand\thesubsection{\Roman{subsection}}

\setcounter{equation}{0}
\renewcommand{\theequation}{S\arabic{equation}}
\renewcommand{\theHequation}{S\arabic{equation}}

In this Supplemental Material we prove the main theorem of the Letter.

\textbf{Theorem.}---For any three-qubit pure state $|\psi\rangle_{ABC}$ with
$\rho_{AB}=\operatorname{Tr}_C(|\psi\rangle_{ABC}\langle\psi|)$,
one has
\begin{equation}\label{eq:thm1-conc}
C(\rho_{AB})+C_a(\rho_{AB})-\sqrt{1-C^2\!\left(|\psi\rangle_{C|AB}\right)}
\le 1.
\end{equation}
Equivalent formulations are
\begin{equation}\label{eq:thm1-cren}
\widetilde{\mathcal{N}}(\rho_{AB})+\widetilde{\mathcal{N}}_a(\rho_{AB})-\sqrt{1-\widetilde{\mathcal{N}}^2\!\left(|\psi\rangle_{C|AB}\right)}
\le 1,
\end{equation}
and
\begin{equation}\label{eq:thm1-neg}
\mathcal{N}(\rho_{AB})+\widetilde{\mathcal{N}}_a(\rho_{AB})-\sqrt{1-\mathcal{N}^2\!\left(|\psi\rangle_{C|AB}\right)}
\le 1.
\end{equation}

\emph{Proof}.---Up to local unitary transformations, any three-qubit pure state can be written in the Ac\'{\i}n canonical form~\cite{1}
\begin{equation}\label{eq:acin}
|\varphi\rangle_{ABC}
=\lambda_{0}|000\rangle+\lambda_{1}e^{i\phi}|100\rangle
+\lambda_{2}|101\rangle+\lambda_{3}|110\rangle+\lambda_{4}|111\rangle,
\end{equation}
where $0\le \phi\le \pi$, $\lambda_s\ge 0$ for $s=0,1,2,3,4$, and $\sum_{s=0}^4\lambda_s^2=1$.

For this parametrization,
\[
C(\rho_{AB})=2\lambda_0\lambda_3,\qquad
C_a(\rho_{AB})=2\lambda_0\sqrt{\lambda_3^2+\lambda_4^2},
\]
and
\begin{equation}\label{eq:Ccab}
C(|\varphi\rangle_{C|AB})
=2\sqrt{\lambda_0^2\lambda_2^2+\lambda_0^2\lambda_4^2+\lambda_1^2\lambda_4^2+\lambda_2^2\lambda_3^2
-2\lambda_1\lambda_2\lambda_3\lambda_4\cos\phi }.
\end{equation}
Define
\[
r:=\sqrt{\lambda_3^2+\lambda_4^2},\qquad
t:=\lambda_0(\lambda_3+r),\qquad
s:=\lambda_1^2+\lambda_2^2=1-\lambda_0^2-r^2,
\]
and
\begin{equation}\label{eq:fdef}
f:=1+\sqrt{1-4A(\phi)}-2t,
\end{equation}
where
\begin{equation}\label{eq:Aphi}
A(\phi):=\lambda_0^2\lambda_2^2+\lambda_0^2\lambda_4^2+\lambda_1^2\lambda_4^2+\lambda_2^2\lambda_3^2
-2\lambda_1\lambda_2\lambda_3\lambda_4\cos\phi .
\end{equation}
It suffices to show that $f\ge 0$ for all admissible parameters.

\medskip
\noindent\textbf{Step 1: Reduction to the worst-case value of $\phi$.}
Since $\lambda_s\ge 0$ and $\cos\phi\in[-1,1]$, the term $-2\lambda_1\lambda_2\lambda_3\lambda_4\cos\phi$ is maximal at $\cos\phi=-1$. Hence
\[
A(\phi)\le A_{\mathrm{wc}}:=\lambda_0^2\lambda_2^2+\lambda_0^2\lambda_4^2+\lambda_1^2\lambda_4^2+\lambda_2^2\lambda_3^2
+2\lambda_1\lambda_2\lambda_3\lambda_4.
\]
Moreover, the function $A\mapsto 1+\sqrt{1-4A}-2t$ decreases on its domain, and therefore
\begin{equation}\label{eq:f-wc}
f \ge 1+\sqrt{1-4A_{\mathrm{wc}}}-2t.
\end{equation}
Thus it is enough to show that the right-hand side of Eq.~(\ref{eq:f-wc}) is nonnegative.

We rewrite
\begin{equation}\label{eq:Awc}
A_{\mathrm{wc}}
=\lambda_0^2(\lambda_2^2+\lambda_4^2)+(\lambda_2\lambda_3+\lambda_1\lambda_4)^2.
\end{equation}

\medskip
\noindent\textbf{Step 2: Maximization at fixed $(\lambda_0,\lambda_3,\lambda_4)$.}
Fix $(\lambda_0,\lambda_3,\lambda_4)$, so that $r$ and $s$ are fixed, and maximize $A_{\mathrm{wc}}$ over $(\lambda_1,\lambda_2)$ subject to $\lambda_1^2+\lambda_2^2=s$.
Let $u:=(\lambda_1,\lambda_2)^{\mathsf T}$. Then
\[
A_{\mathrm{wc}}=\lambda_0^2\lambda_4^2+u^{\mathsf T}Nu,
\qquad
N:=\begin{pmatrix}
\lambda_4^2 & \lambda_3\lambda_4\\
\lambda_3\lambda_4 & \lambda_3^2+\lambda_0^2
\end{pmatrix}.
\]
The matrix $N$ is positive semidefinite because
\[
\operatorname{Tr}(N)=\lambda_0^2+\lambda_3^2+\lambda_4^2\ge 0,
\qquad
\det(N)=\lambda_0^2\lambda_4^2\ge 0.
\]
By the Rayleigh quotient,
\[
u^{\mathsf T}Nu \le \lambda_{\max}(N)\,\|u\|^2=\lambda_{\max}(N)\,s,
\]
and hence
\begin{equation}\label{eq:Amax-def}
A_{\mathrm{wc}}\le A_{\max}:=\lambda_0^2\lambda_4^2+s\,\lambda_{\max}(N).
\end{equation}
Because the right-hand side of Eq.~(\ref{eq:f-wc}) decreases with $A_{\mathrm{wc}}$,
\begin{equation}\label{eq:reduce}
f \ge 1+\sqrt{1-4A_{\max}}-2t.
\end{equation}

To evaluate $\lambda_{\max}(N)$, define
\[
U:=\lambda_0^2+r^2=\lambda_0^2+\lambda_3^2+\lambda_4^2,
\qquad
\Delta:=\sqrt{U^2-4\lambda_0^2\lambda_4^2}.
\]
Since $\operatorname{Tr}(N)=U$ and $\det(N)=\lambda_0^2\lambda_4^2$, the eigenvalues of $N$ are $(U\pm\Delta)/2$, so
\begin{equation}\label{eq:lmax}
\lambda_{\max}(N)=\frac{U+\Delta}{2}.
\end{equation}
Using $s=1-U$, Eq.~(\ref{eq:Amax-def}) becomes
\begin{equation}\label{eq:Amax}
A_{\max}=\lambda_0^2\lambda_4^2+s\,\frac{U+\Delta}{2},\qquad s=1-U.
\end{equation}

\medskip
\noindent\textbf{Step 3: A key identity.}
A direct simplification of Eq.~(\ref{eq:Amax}), together with $\Delta^2=U^2-4\lambda_0^2\lambda_4^2$, yields
\begin{equation}\label{eq:keyid}
1-4A_{\max}=(\Delta-s)^2\ge 0.
\end{equation}
Therefore,
\[
\sqrt{1-4A_{\max}}=|\Delta-s|\ge \Delta-s.
\]
Substituting this into Eq.~(\ref{eq:reduce}) and using $1-s=U$, we obtain
\begin{equation}\label{eq:lower}
f \ge U+\Delta-2t.
\end{equation}

\medskip
\noindent\textbf{Step 4: Final estimate.}

\smallskip
\noindent\emph{Case 1: $t\le \tfrac12$.}
Since $\sqrt{1-4A_{\max}}\ge 0$, Eq.~(\ref{eq:reduce}) immediately gives
\[
f \ge 1-2t \ge 0.
\]

\smallskip
\noindent\emph{Case 2: $t> \tfrac12$.}
Since $U=\lambda_0^2+r^2\le \sum_{k=0}^4\lambda_k^2=1$, we have $2t>1\ge U$, and hence $2t-U\ge 0$.
By Eq.~(\ref{eq:lower}), it is enough to prove
\begin{equation}\label{eq:target}
U+\Delta \ge 2t = 2\lambda_0(\lambda_3+r).
\end{equation}
Equivalently,
\[
\Delta \ge 2\lambda_0(\lambda_3+r)-U.
\]
The right-hand side is nonnegative, so squaring preserves the inequality:
\[
\Delta^2 \ge \bigl(2\lambda_0(\lambda_3+r)-U\bigr)^2.
\]
Using $\Delta^2=U^2-4\lambda_0^2\lambda_4^2$ and canceling $U^2$, we obtain
\[
-4\lambda_0^2\lambda_4^2 \ge 4\lambda_0^2(\lambda_3+r)^2-4\lambda_0(\lambda_3+r)U.
\]
Rearranging gives
\[
4\lambda_0(\lambda_3+r)\bigl(U-\lambda_0(\lambda_3+r)\bigr)\ge 4\lambda_0^2\lambda_4^2.
\]
Because $t>1/2$ implies $\lambda_0(\lambda_3+r)>0$, division by $4\lambda_0(\lambda_3+r)$ is legitimate. Now
\[
U-\lambda_0(\lambda_3+r)
=\lambda_0^2+r^2-\lambda_0\lambda_3-\lambda_0r
=(\lambda_0-r)^2+\lambda_0(r-\lambda_3),
\]
and
\[
\lambda_4^2=r^2-\lambda_3^2=(r-\lambda_3)(r+\lambda_3).
\]
Hence the difference between the two sides is
\begin{align*}
&4\lambda_0(\lambda_3+r)\bigl(U-\lambda_0(\lambda_3+r)\bigr)-4\lambda_0^2\lambda_4^2\\
&=4\lambda_0(\lambda_3+r)\Big[(\lambda_0-r)^2+\lambda_0(r-\lambda_3)\Big]
-4\lambda_0^2(r-\lambda_3)(r+\lambda_3)\\
&=4\lambda_0(r+\lambda_3)(\lambda_0-r)^2 \ge 0.
\end{align*}
Thus Eq.~(\ref{eq:target}) holds, and therefore $f\ge 0$ also in Case 2.

Combining the two cases, we conclude that $f\ge 0$ for all admissible parameters, which proves Eq.~(\ref{eq:thm1-conc}).

For the reduced two-qubit state $\rho_{AB}$ of a three-qubit pure state,
\begin{equation}\label{eq:twoqubit-cren-conc}
\widetilde{\mathcal{N}}(\rho_{AB})=C(\rho_{AB}),
\qquad
\widetilde{\mathcal{N}}_a(\rho_{AB})=C_a(\rho_{AB}).
\end{equation}
Moreover, for the pure bipartition $C|AB$,
\begin{equation}\label{eq:pure-neg-conc-cab}
\mathcal{N}\!\left(|\psi\rangle_{C|AB}\right)
=
C\!\left(|\psi\rangle_{C|AB}\right).
\end{equation}
Substituting Eqs.~(\ref{eq:twoqubit-cren-conc}) and (\ref{eq:pure-neg-conc-cab}) into Eq.~(\ref{eq:thm1-conc}) yields Eq.~(\ref{eq:thm1-cren}). Finally, since
\begin{equation}\label{eq:neg-less-cren}
\mathcal{N}(\rho_{AB})\le \widetilde{\mathcal{N}}(\rho_{AB}),
\end{equation}
Eq.~(\ref{eq:thm1-neg}) follows immediately.
\hfill$\square$

\end{document}